\documentclass[a4paper,11pt]{article}
\usepackage{physics}
\usepackage{amsmath,bm}
\usepackage{graphicx}
\usepackage{pos}
\usepackage{bbold}

\def\lsim{\raise0.3ex\hbox{$<$\kern-0.75em\raise-1.1ex\hbox{$\sim$}}}
\def\gsim{\raise0.3ex\hbox{$>$\kern-0.75em\raise-1.1ex\hbox{$\sim$}}}

\title{Efficient simulation of quarkonium master equation beyond the dipole approximation}

\author*[a,b,c]{Jorge M. Mtz-Vera}
\emailAdd{jorgemanuel.martinezvera@unito.it}

\author[a]{Andrea Beraudo}
\emailAdd{beraudo@to.infn.it}

\author[c]{Miguel Ángel Escobedo}
\emailAdd{miguel.a.escobedo@fqa.ub.edu}

\author[a,b]{Paolo Parotto}
\emailAdd{paolo.parotto@unito.it}

\author{Michael Strickland}
\emailAdd{michael.strickland@gmail.com}

\affiliation[a]{INFN - Sezione di Torino, via Pietro Giuria 1, I-10125 Torino, Italy}

\affiliation[b]{Dipartimento di Fisica, Università di Torino and INFN Torino, Via P. Giuria 1, I-10125 Torino, Italy}

\affiliation[c]{Institut de Ciències del Cosmos, Universitat de Barcelona, Martí i Franquès 1, 08028 Barcelona, Catalonia}

\date{\today}
\abstract{QTRAJ is a computer code that simulates the propagation of quarkonium in the quark-gluon plasma (QGP) based on the quantum-trajectory algorithm. This algorithm solves a master equation in which the quarkonium is treated as an open quantum system (OQS).
A major advantage of this approach is that it turns a 3D spatial evolution for a density matrix into a 1D Schr\"odinger equation for a wavefunction with a non-hermitian Hamiltonian, drastically reducing the computational cost.
So far, the interaction implemented in the master equation was obtained within the framework of potential non-relativistic QCD (pNRQCD), and restricted to the regime $rT \ll 1$, where $r$ is the size of the color dipole and $T$ is the temperature. In the environment produced in heavy-ion collisions (HIC's) this limit is accurate for $\Upsilon(1S)$, but the applicability to other quarkonium states is dubious.
In the present study we generalize the above approach, extending it to the regime $rT\!\sim\! 1$ in the one-gluon exchange approximation, with proper Hard Thermal Loop (HTL) resummation of medium effects. This is done by implementing new jump operators connecting different color states of the $Q\bar Q$ pair and expanding them in plane waves, giving rise to a variation of the algorithm present in QTRAJ 1.0. Here we
 provide an overview of this approach comparing the $rT \ll 1$ and $rT\sim 1$ cases, and we discuss prospects for phenomenological application to excited states of bottomonium.}

\FullConference{10th International Conference on Quarks and Nuclear Physics (QNP2024)\\
8-12 July, 2024\\
Barcelona, Spain\\}

\begin{document}
\maketitle

\section{Introduction: in-medium quarkonium as an Open Quantum System}
 Quarkonium in the QGP can be treated as a small system ($Q$, described by the Hamiltonian $H_Q$) coupled to a big environment ($E$ described by the Hamiltonian $H_E$ and made of light degrees of freedom) via the interaction $H_I$. In full generality the total Hamiltonian is thus given by:
\begin{equation}
H_T=H_Q\otimes  \mathbb{1}_E+\mathbb{1}_Q\otimes H_E+H_I.
\end{equation}
 The full $Q+E$ evolution is described by the von Neumann equation for the total density matrix $\rho_T$. However, within the Open Quantum System (OQS) approach~\cite{Breuer:2002pc}, one can conveniently evaluate the expectation value of any observable involving only the small system $Q$ by employing the reduced density matrix $\rho_Q$, obtained by taking the trace over the environmental degrees of freedom of $\rho_T$, i.e. $\rho_Q\equiv{\rm Tr}_E(\rho_T)$. 

In the case of quarkonium in a quark-gluon plasma (QGP), different timescales characterize the problem. There is the system evolution timescale $\tau_Q\!\sim\! 1/\Delta E$, where $\Delta E$ is the energy gap between excited states (if $\Delta E$ were the binding energy in a Coulomb potential $\tau_Q$ would be the orbital period); there is the equilibration time of the light environment degrees of freedom $\tau_E\!\sim\!1/T$, where $T$ is the temperature; finally, there is the system equilibration time $\tau_R \sim M_Q/T^2$, where $M_Q$ is the heavy quark mass. In order to obtain a tractable form of the evolution equation for the reduced density matrix, approximations are performed: the Born-Markov approximation (weak coupling with the medium and absence of memory effects) and the Born-Oppenheimer approximation (very fast accommodation of the light degrees of freedom after perturbations induced by the heavy ones). In terms of the characteristic timescales of the system this corresponds to $\tau_E \ll \tau_Q$ and $\tau_E \ll \tau_R$. This regime is called the Quantum Brownian Motion regime (QBM), in which the evolution equation for the reduced density matrix takes the following Lindblad form:
\begin{equation}
        \frac{d\rho_Q}{dt} = -i \left[H'_Q, \rho_Q \right] + \sum_{k} \left( C_k \rho_Q C_k^\dagger - \frac{1}{2} \{ C_k^\dagger C_k, \rho_Q \} \right),\label{eq:Lindblad}
\end{equation}
where $C_k$ are the so-called {\em collapse} (or {\em jump}) operators allowing transitions between two different states, and $H'_Q$ is a hermitian Hamiltonian which includes medium corrections to the heavy quark self-energy and $Q\overline Q$ potential (Debye screening in our case).
In the present study both the jump operators and the medium corrections to the real part of the Hamiltonian are evaluated in the Hard Thermal Loop (HTL) approximation \cite{Beraudo:2007ky}. In principle the latter relies on a scale separation between the typical light-parton momentum $\sim\!\pi T$ and the Debye mass $m_D\!\sim\! gT$. In practice, at the experimentally accessible temperatures the coupling is not so weak, $g\!\sim\!2$, and the scale separation is only marginally satisfied. Nevertheless here we adopt this theoretical framework, treating the Debye mass as a free parameter. This allows one to capture a rich set of medium effects, not yet fully explored in the literature.

For our study it is convenient to write the system density matrix \cite{Brambilla:2017zei}
in the singlet-octet basis, where it has the diagonal structure
\begin{equation}
\rho_Q(t) = \rho_s(t) \ket{s}\bra{s} + \bar{\rho}_o(t) \sum_c \ket{o^c}\bra{o^c}\equiv \rho_s(t) \ket{s}\bra{s} + {\rho}_o(t) \ket{o}\bra{o}\,,
\end{equation}
 having taken an average over all octet states, namely (we are assuming that there is no preferred direction in color space \cite{Escobedo:2019gzn}):
\begin{equation}
    \rho_o\equiv (N_c^2-1)\bar\rho_o\,,\quad \ket{o}\bra{o}\equiv\frac{1}{N_c^2-1}\sum_c \ket{o^c}\bra{o^c}\,.
\end{equation}

Asymptotically, the evolution driven by Eq.~(\ref{eq:Lindblad}) will lead the system to a maximum entropy state, with $\rho_s=1/9$ and $\rho_o=8/9$ \cite{Blaizot:2018oev}. The code performance, which will be described in the following, can be tested by using this important consistency check on its output.

\section{The method: quantum trajectories.}\label{sec:method}
 Obtaining a direct numerical solution from the Lindblad equation would be extremely inefficient.
Studying the image of the evolution of $\rho_Q$ in coordinate space requires the implementation of Eq.~(\ref{eq:Lindblad}) in a spatial lattice of $N^3$ nodes. Even though we can focus just on the relative motion of the $Q\bar Q$ pair, this still entails the evaluation, at each time-step, of $N^3\times N^3$ matrix elements. Instead, it is far more convenient to use an equivalent formulation of the problem in terms of a stochastic evolution of the wavefunction, driven by a non-hermitian Hamiltonian. This is the idea at the basis of the quantum trajectories algorithm (which is a Monte Carlo Wavefunction method \cite{KORNYIK201988}), allowing one to halve the dimensionality of the problem~\cite{Daley:2014fha}. 

Furthermore, one can assume that at each time step the $Q\overline Q$ wavefunction is a well defined eigenstate of the angular orbital momentum (the in-medium $Q\overline Q$ potential is central if the pair does not move too fast), which amounts to assuming a diagonal form for the reduced density matrix not only in color, but also in the space of the $\bm{L}^2$ eigenvalues. Hence, employing a spherical-harmonics basis allows one to reduce the problem to a one-dimensional evolution, involving only the relative radial coordinate of the pair. 

The quantum trajectories algorithm revolves around two pillars:
\begin{enumerate}
    \item Simulating independent single trajectories of quarkonium, in which the pair wavefunction is evolved stochastically. At the beginning and at the end of the evolution the in-medium wavefunction is projected onto the physical vacuum-states of quarkonium. The final result is obtained by averaging over a large sample of independent trajectories. Due to this independence, the code can be trivially parallelized.
    \item The rearrangement of the Lindblad equation into a form involving two different terms: a deterministic part corresponding to a Schrödinger-like evolution driven by a non-hermitian Hamiltonian; a stochastic part, with the so-called quantum restitution term, which will induce sudden transitions of the pair to a different state.
    \end{enumerate}
    
The aformentioned non-hermitian Hamiltonian is given by:
\begin{equation} 
H_{\text{eff}} \equiv H'_Q - \frac{i}{2} \sum_n \int_{\bm q}  C_{\bm q}^{n\dagger} C_{\bm q}^n\equiv  H'_Q - \frac{i}{2} \sum_n \int_{\bm q} \Gamma_{\bm q}^n\equiv H'_Q - \frac{i}{2}\Gamma.\label{eq:Heff}
\end{equation}
The latter is the analogous of the optical potential employed in nuclear physics, with an imaginary part accounting for the decrease of the norm of the state due to transitions to any possible final state, in this case identified by the color ($n$) and linear-momentum ($\bm q$) exchange involved in the jump. The Lindblad equation takes then the form:
\begin{equation}
{\frac{d \rho_Q}{dt} = - i [H_{\text{eff}},\rho_Q]} + \sum_{n} \int_{\bm q} C_{\bm q}^n\rho_Q C_{\bm q}^{n\dagger}.\label{eq:rhoeff}
\end{equation}

We now briefly describe the quantum-trajectory algorithm in full generality, independently from the specific form of the effective Hamiltonian and jump operators. 
\begin{enumerate}
    \item A random number, $r_0 \in [0,1]$, is picked at the beginning. Then, a Schrödinger-like evolution of the radial wavefunction is performed by employing $H_{\text{eff}}$. The norm of the state decreases, due to the non-hermiticity of $H_{\text{eff}}$.
    
    \item At every time step, the code verifies if the norm of the state of the system $Q$ is still greater than $r_0$. If so, the simulation continues with another time step.
    
    \item If the norm goes below $r_0$, the deterministic evolution stops; we say that a jump has been triggered. By extracting some further random numbers, we select the final state of the transition, based on the corresponding exclusive decay width.
    
    \item Having selected a specific transition, we apply the corresponding Lindblad operator to the initial wavefunction to obtain the final state, whose norm is set to 1 and the cycle starts again.
\end{enumerate}

We now display the form assumed by the above operators when the system is a $Q\overline Q$ pair and its interaction with the QGP is described within the HTL approximation \cite{Akamatsu:2020ypb} via a one-gluon exchange. In this case the in-medium hermitian Hamiltonian takes the form
\begin{equation}
    H'_Q\equiv
    \begin{pmatrix}
        H'_s & 0\\
        0 & H'_o
    \end{pmatrix}
%    diag(H'_s, H'_o)
    \equiv
    -C_F\alpha_s m_D \mathbb{1} \hspace{2mm}
    + \hspace{2mm} \frac{\alpha_s}{r}e^{-m_Dr}
\begin{pmatrix}
-C_F & 0\\
0 & 1/2N_c
\end{pmatrix}\,,
    %\cdot diag (
    %    -C_F ,\frac{1}{2N_c}),
\label{eq:ReVeff}
\end{equation}
while the jump operators read
\begin{equation}
C_{\bm q}^{0}\equiv\begin{pmatrix}
    0 & (1/\sqrt{2N_c})L_{\bm q}\\
    \sqrt{C_F}L_{\bm q} & 0
\end{pmatrix},\quad
C_{\bm q}^{1}\equiv\sqrt{\frac{N_c^2-4}{4N_c}}\begin{pmatrix}
0 & 0\\
0 & L_{\bm q}
 \end{pmatrix},\quad
C_{\bm q}^{2}\equiv\sqrt{\frac{N_c}{4}}\begin{pmatrix}
0 & 0\\
0 & \overline{L}_{\bm q}
 \end{pmatrix} \,,\label{eq:HTLjump}
\end{equation}
where
\begin{equation}
    L_{\bm q}\equiv g\sqrt{\Delta^>(\omega=0,q)}S_{\bm q\cdot \bm{\hat r}}\quad{\rm and}\quad
    \overline L_{\bm q}\equiv g\sqrt{\Delta^>(\omega=0,q)}C_{\bm q\cdot \bm{\hat r}}\,.
\end{equation}
In the above $\Delta^>(\omega)$ if the Fourier transform of the in-medium HTL propagator $\langle A^0(t)A^0(0)\rangle$, while 
\begin{equation}
S_{\bm q\cdot \bm{\hat r}}\equiv 2\sin(\bm q \cdot \bm{\hat r}/2),\quad
C_{\bm q \cdot \bm{\hat r}}\equiv 2\cos(\bm q \cdot \bm{\hat r}/2)\,.
\end{equation}
The jump operators in Eq.~(\ref{eq:HTLjump}) can be obtained by expressing the results of Ref.~\cite{Blaizot:2018oev} in the present Lindblad form, more suited to a numerical implementation. $C_{\bm q}^0$ describes parity-changing $s\leftrightarrow o$ transitions; $C_{\bm q}^1$ and $C_{\bm q}^2$ describe parity-changing and parity-conserving $o\to o$ transitions, respectively. They differ from the ones employed in the original QTRAJ 1.0 code~\cite{Omar:2021kra, Brambilla:2022ynh} both for not being limited to the dipole approximation, applicable only to a small-size pair, and for the appearance of the novel $C_{\bm q}^2$ parity-conserving $o\to o$ jump operator. This represents the major theoretical contribution of our study which, after its numerical implementation, will allow the application of the QTRAJ code to more general situations. Notice that interactions with the medium occur only through the exchange of space-like gluons within this setup, while direct gluo-dissociation processes are neglected.

\section{New selection rules}
The stochastic part of our algorithm, simulating sudden transitions induced by collisions with medium particles, is structured around selection rules that specify which jump operator has to be used to obtain the final state.

In the soon-to-be-released QTRAJ 1.1 there are four random choices to perform: 1) the color channel, 2) the transferred linear momentum, 3) the angular momentum of the exchanged gluon and 4) the angular momentum of the final state. These choices are made sequentially. Although steps 2), 3) can be taken in any order, we choose this one because of the dramatic reduction in computation time.

Having selected the transition, the final state will be
\begin{equation}
    \ket{\psi_{\rm new}} = \frac{C^n_{\bm q} \ket{\psi_{\rm old}}}{\sqrt{\bra{\psi_{\rm old}}\Gamma^n_{\bm q}\ket{\psi_{\rm old}}}}\,,\quad{\rm where}\quad\Gamma^n_{\bm q}=C^{n\dagger}_{\bm q}C^n_{\bm q}\;\longrightarrow\; \bra{\psi_{\rm new}}\ket{\psi_{\rm new}}=1\,.
\end{equation}

Parity-changing/conserving transitions involve widths proportional to 
\begin{equation}
    \Gamma_-(r)=
    2g^2\int_{\bm q}\Delta^>(0,q)\sin^2\left({\bm q\!\cdot \bm r}/2\right)\quad{\rm and}\quad
\Gamma_+(r)=
    2g^2\int_{\bm q}\Delta^>(0,q)\cos^2\left({\bm q\!\cdot \bm r}/2\right)\,,\label{eq:gammapm}    
\end{equation}
respectively, properly convoluted with the squared wavefunction of the initial state.
\subsection{Color channel}
 Any single-gluon exchange transition of a color-singlet $Q\overline Q$ pair can only lead to an octet, hence $p_{s\to o}\!=\!1$.
However, starting from an octet state there are several possibilities, its total width being given by the sum $\Gamma_o=\Gamma_o^0+\Gamma_o^1+\Gamma_o^2$. Each partial width $\Gamma_o^i$ ($i=0,1,2$) arises from the corresponding jump operator in Eq.~(\ref{eq:HTLjump}), describing $o\to s$ transitions ($C_{\bm q}^{0}$), parity-changing $o\to o$ transitions ($C_{\bm q}^{1}$) and parity conserving $o\to o$ transitions ($C_{\bm q}^{2}$). The conditional probability used to randomly select the specific jump is given by the ratio $p_{o\to i}=\Gamma_o^i/\Gamma_o.$

\subsection{Linear momentum transfer}

Depending on whether one is dealing with a parity-changing (-) or parity conserving (+) transition, the modulus of the linear momentum exchanged $q$ is randomly extracted from the distribution
\begin{equation}
\frac{d \Gamma_{\mp}}{dq}\sim \int_0^\infty dr |u_l(r)|^2
\frac{2 q\, m_D^2}{(q^2+m_D^2)^2}
\left[1\mp\frac{\sin(q r)}{qr}\right]
\,,\label{eq:dGammambdq} 
\end{equation}
whose specific expression arises from the HTL result for the longitudinal gluon propagator $\Delta^>(0,q)$ \cite{Beraudo:2007ky}. In the above $u_l(r)$ is the radial wavefunction of the state, corresponding to a given eigenvalue $l$ of the orbital angular momentum.

\subsection{Angular momentum eigenvalue of the virtual gluon}
 Having extracted $q$, which will be kept fixed, one can select the orbital angular momentum exchanged in the collision. This can be done by expanding the integrand in Eq.~(\ref{eq:dGammambdq}) in spherical Bessel functions.

Each contribution to the expansion can be considered as the conditional probability of exchanging, at fixed $q$, a given angular momentum $l_g\!=\!2t\!+\!1\,{\rm or}\,2t$ (with $t=0,1,\dots\infty$), which can then be randomly selected. One has:
\begin{equation}
p_-(t|q)\!\sim\!(4t+3)\!\int_0^\infty dr |u_l(r)|^2\,|j_{2t+1}(qr/2)|^2\,,\quad
p_+(t|q)\!\sim\! (4t+1)\!\int_0^\infty dr |u_l(r)|^2\,|j_{2t}(qr/2)|^2\,.
\end{equation}
The overall factors correspond to the degeneracy from the $2l_g\!+\!1$ projections along the quantization axis. If the wavefunction of the pair is not too broad only the first few terms of the above expansion are sufficient to saturate the sum.

\subsection{Angular momentum eigenvalue of the final state}
 Having fixed $l_g$ in step 3 only a limited number of final angular momentum states $\ket{l_f,m_f}$ are accessible to the pair. These can be obtained from the corresponding Clebsch-Gordan coefficients.
 Summing over all possible projections along the quantization axis (to which we are not interested), one
 gets the following conditional probability for a given final angular momentum state
\begin{equation}
    p(l_i \rightarrow l_f |l_g) = \abs{\bra{l_i,0; l_g ,0}\ket{l_f,0}}^2,
\end{equation}
with $l_g = 2t+1$ and $l_g = 2t$ for parity-changing and parity-conserving interactions, respectively.

\section{Preliminary results and discussion}
\begin{figure}
\centering
\includegraphics[width=0.495\textwidth]{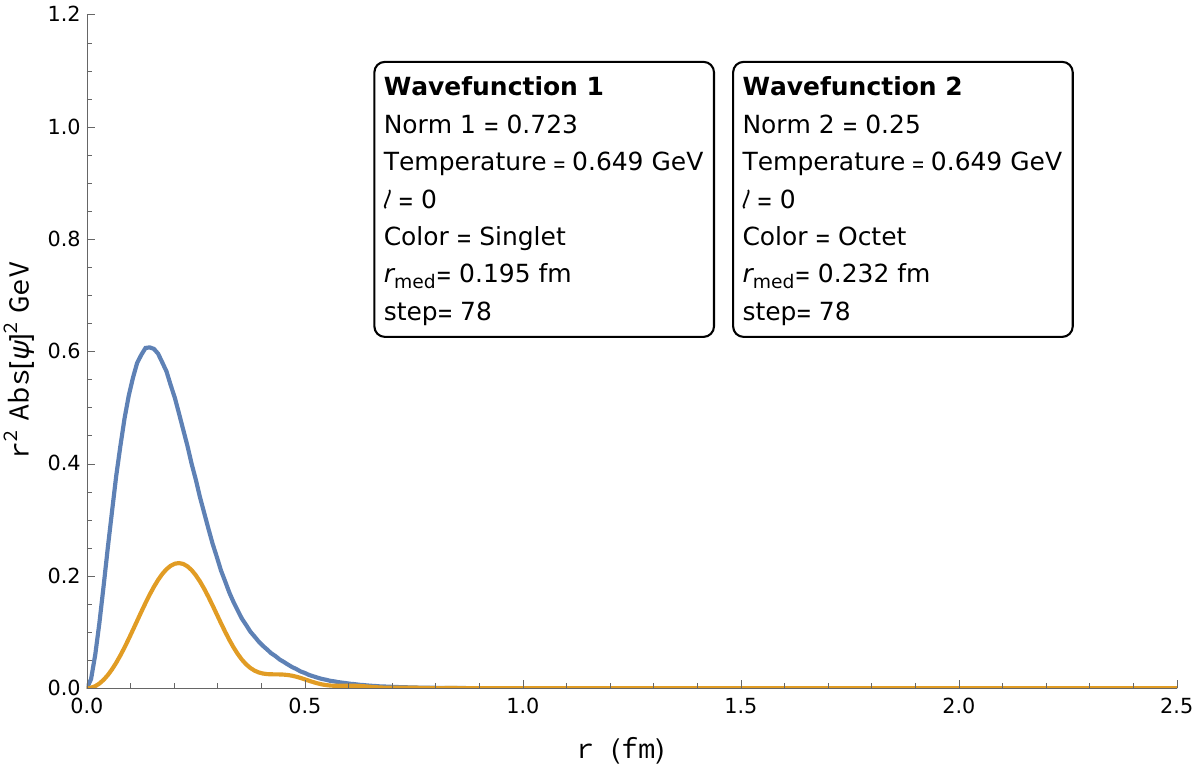}
\includegraphics[width=0.495\textwidth]{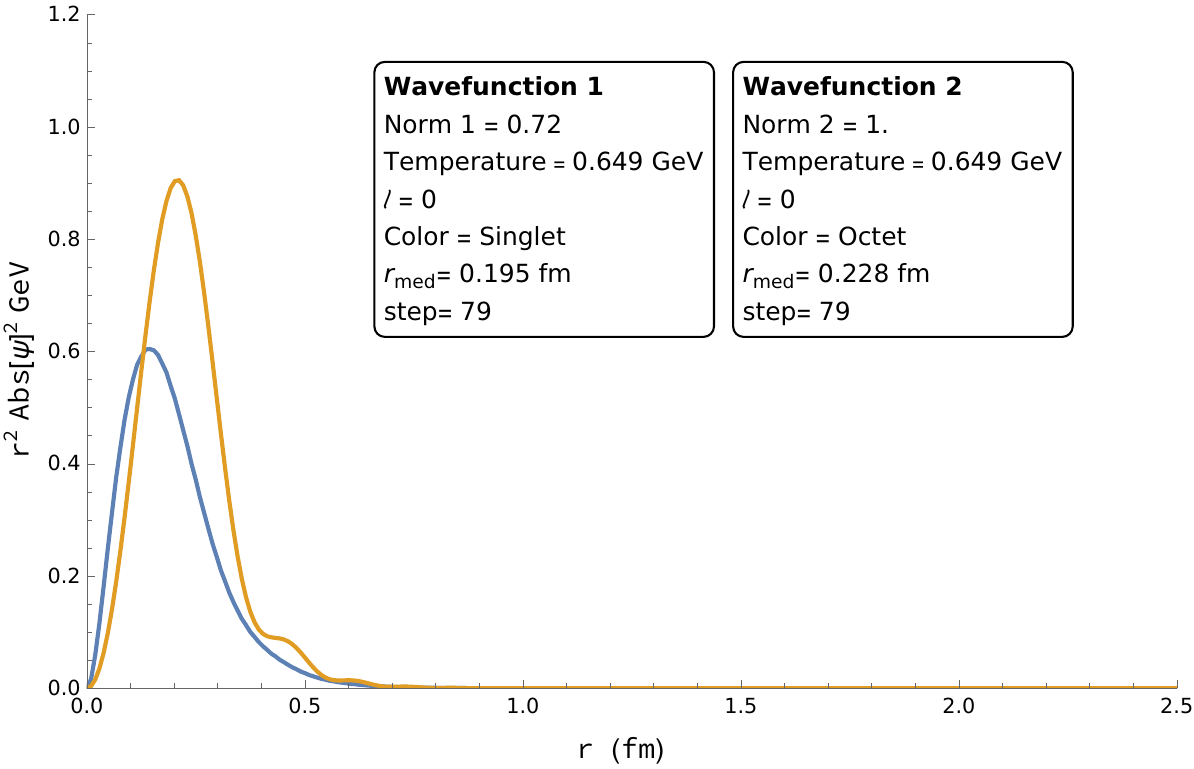}
\caption{Snapshot of $Q\overline Q$ color-singlet (blue) and octet (orange) wavefunctions before (left panel) and after (right panel) an $ o\to o$ transition for the second pair.}
\label{fig:wf}
\end{figure}
We show an example of the results obtained with QTRAJ 1.1 in Fig.~\ref{fig:wf}. We display two independent trajectories for a $Q\overline Q$ pair, before and after a transition for one of the pairs. They were initialized to a color-singlet and a color-octet state respectively. Both pairs start their evolution in a medium at $T=700$~MeV with the same radial wavefunction of average radius $\langle r\rangle=0.2$ fm. Both pairs propagate in a medium undergoing a Bjorken expansion, with a power-law cooling $T\sim t^{-1/3}$. One can see that, while the singlet state -- in which the $Q\overline Q$ interaction is attractive -- remains quite compact, in the octet channel the pair wavefunction gets broader, due to the repulsive potential. 
A color-singlet $Q\overline Q$ pair, as long as $r\,\lsim\, r_D\!\equiv\! 1/m_D$, undergoes very rare collisions with the medium. On the other hand, in the color-octet channel even a very small pair is seen by the medium as a charged object, an effective gluon. Hence for the imaginary part of the potential one has $\Gamma_s <\Gamma_o$, entailing that the norm of the state (representing in our approach the fraction of pairs which have not suffered any collision) decays faster in the octet channel. On the other hand, once a quantum jump has taken place the norm is set back to unity, as shown in the right panel. For more results and consistency checks of our approach we refer the reader to a forthcoming publication.

\acknowledgments
A.B. acknowledges financial support by MUR within the Prin$\_$2022sm5yas project. The work of MAE and JMMV have been supported by the Maria de Maetzu excellence program under project CEX2019-000918-M, and by project PID2022-136224NB-C21 funded by MCIN/AEI/10.13039/ 501100011033, and MAE also by grant 2021-SGR-249 of Generalitat de Catalunya.

\bibliographystyle{JHEP}
\bibliography{ProceedingsBIB}

\end{document}